\documentclass[prl,superscriptaddress,twocolumn,aps,showpacs]{revtex4}
\usepackage{amssymb}
\usepackage{graphicx}
\usepackage{longtable}
\usepackage{amsmath}
\usepackage{amsfonts}
\usepackage{color}
\begin{document}
\author{E.~Gorelov}
\affiliation{Institut f\"ur Festk\"orperforschung and Institute for Advanced Simulation, Forschungszentrum J\"ulich, 
             52425 J\"ulich, Germany}
\author{M.~Karolak}
\author{T.~O.~Wehling}
\author{F.~Lechermann}
\author{A.I.~Lichtenstein}
\affiliation{ 
I. Institut f{\"u}r Theoretische Physik, Universit{\"a}t Hamburg, 
Jungiusstra{\ss}e 9, D-20355 Hamburg, Germany}
\author{E.~Pavarini}
\affiliation{Institut f\"ur Festk\"orperforschung and Institute for Advanced Simulation, Forschungszentrum J\"ulich, 
             52425 J\"ulich, Germany}
\date{\today }
\title{Nature of the Mott transition in Ca$_2$RuO$_4$}
\begin{abstract}
  We study the origin of the temperature-induced  Mott transition in Ca$_2$RuO$_4$. As a method we use the 
  local-density approximation+dynamical mean-field theory. 
  We show the following. {\em (i)} The Mott transition is driven by the change in 
 structure from long to short {\bf c}-axis layered perovskite (L-Pbca $\to$ S-Pbca); it
 occurs together with orbital order, which follows, rather than produces, the structural transition.
 {\em (ii)} In the metallic L-Pbca phase the orbital polarization is $\sim 0$.  {\em (iii)} In the  insulating S-Pbca phase  the lower energy orbital,
 $\sim xy$, is full.
 {\em (iv)} The spin-flip and pair-hopping Coulomb terms reduce the effective masses in the metallic phase. 
 Our results indicate that a similar scenario applies to Ca$_{2-x}$Sr$_x$RuO$_4$ $(x\le0.2)$. 
 In the metallic $x\le 0.5$ structures electrons are progressively transferred to the $xz/yz$ bands
 with increasing $x$, however we find no orbital-selective Mott transition down to $\sim$~300~K. 
 \end{abstract}
\maketitle

The layered perovskite Ca$_2$RuO$_4$ (Ru 4$d^4$, $t_{2g}^4e_g^0$) undergoes a paramagnetic metal-paramagnetic insulator transition (MIT) at 
$T_{\rm {MIT}}\sim 360$~K \cite{mit}.
A similar insulator-to-metal transition happens also by application of a  modest ($\sim 0.5$~GPa) pressure \cite{pressure}
and finally when Ca is partially substituted by Sr (Ca$_{2-x}$Sr$_{x}$RuO$_4$, $x \leq 0.2$) \cite{str2,ca2x}. 
The nature of these transitions, in particular across $x=0.2$, has been debated for  a decade \cite{OSMT,Fang,Neupane,Shimo,rxr,gap,xas,xas2,liebsch}. 
While it is clear that a Mott-type mechanism makes the 2/3-filled $t_{2g}$ bands insulating,
two opposite scenarios, with different orbital occupations $n=(n_{xy},n_{xz}+n_{yz})$ and
polarizations $p\equiv n_{xy}-(n_{xz}+n_{yz})/2$, have been suggested.  
In the first, only the $xy$ band becomes metallic, i.e. the transition is orbital-selective (OSMT) \cite{OSMT}; 
$n$ and $p$ jump from $(2,2)$ and $1$ in the insulator 
to $(1,3)$ and $-1/2$ in the metal. In the second, there is a single Mott transition, assisted by the crystal-field splitting $\Delta=\epsilon_{xz/yz}-\epsilon_{xy}>0$ \cite{liebsch}, similar to the case of 3$d^1$ perovskites \cite{eva}; $p>0$ in all phases.
To date the issue remains open.
Recently, for $x=0.2$ a novel ($xy$ insulating, $n_{xy}=1.5$ and  $p=1/4$) OSMT was inferred from angle-resolved photoemission (ARPES)  experiments \cite{Neupane}, but other ARPES data show three metallic bands and no OSMT \cite{Shimo}. 

\begin{figure}
\center
\rotatebox {0}{\includegraphics [width=0.32\textwidth]{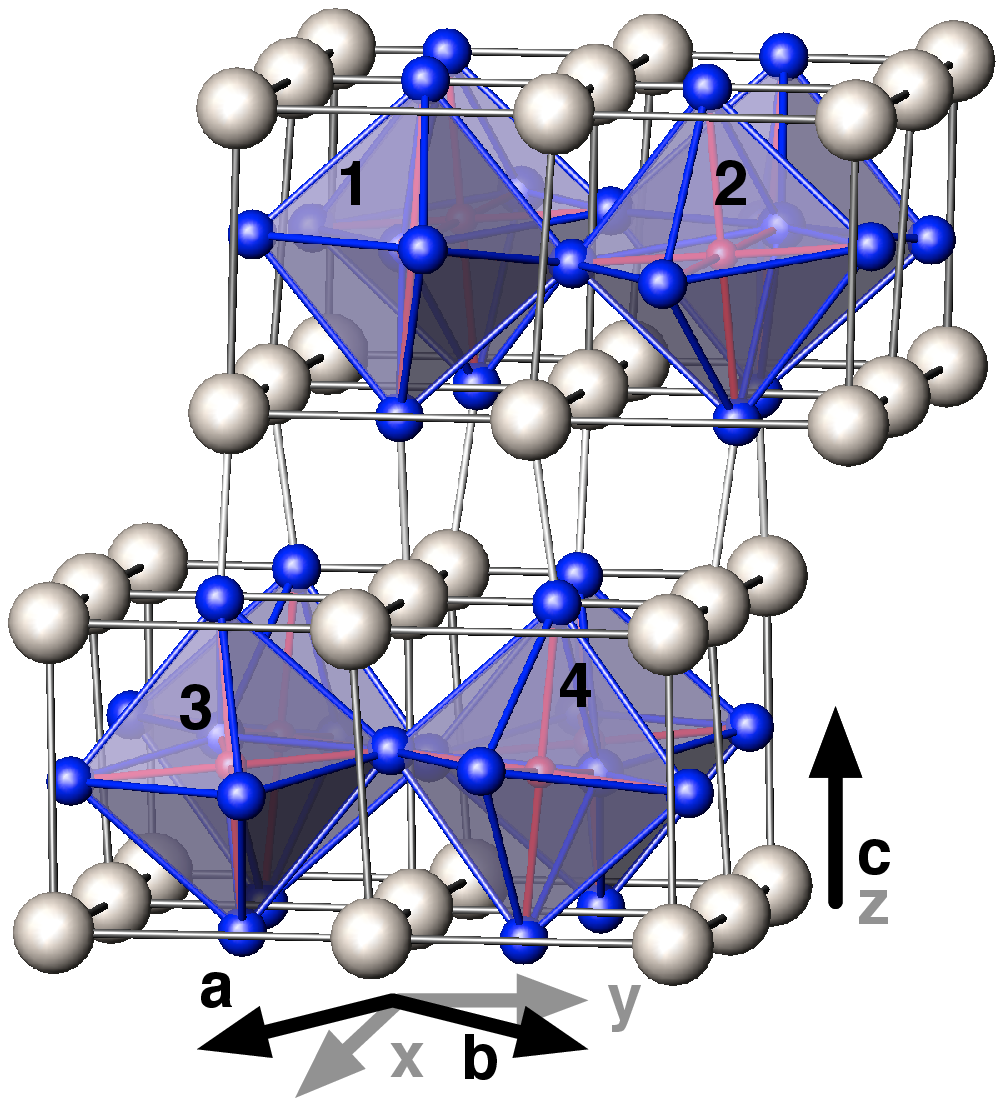}}
\caption{(Color online) \label{crys-str}
Ca$_2$RuO$_4$ L-Pbca \cite{str1,str2}. The primitive cell is orthorhombic with 4 formula units; $ {\bf x}\sim ({\bf a}+{\bf b})/2 $,
${\bf y}\sim({\bf b}-{\bf a})/2$, ${\bf z}={\bf c}$ are the pseudotetragonal axes. 
Ru sites $i$ at ${\bf T}_i$ ($i=2,3,4$) are equivalent to site 1 at ${\bf T}_1$, with operations 
${\bf a}\to -{\bf a}$  ($i=2$), ${\bf c} \to -{\bf c}$ ($i=3$), $ {\bf b} \to -{\bf b}$ ($i=4$) and ${\bf T}_i \to {\bf T}_1$. 
In the S-Pbca structure the tilting angle  is about twice  as large,
while the rotation angle is slightly smaller.
}
\end{figure} 

Ca$_2$RuO$_4$ is  made of RuO$_2$ layers built up of 
corner-sharing RuO$_6$ octhahedra (space group Pbca \cite{str2,str1}). 
This structure (Fig.~\ref{crys-str}) combines a rotation of the octahedra around the ${\bf c}$ axis 
with a tilt around the ${\bf b}$ axis.  It is similar to that of the  tetragonal unconventional 
superconductor Sr$_2$RuO$_4$; the corresponding pseudo-tetragonal 
axes {\bf x}, {\bf y} and {\bf z} are shown in Fig.~\ref{crys-str}. 
The structure of Ca$_2$RuO$_4$ is characterized by a long {\bf c} axis (L-Pbca)
above $T_S\sim356~$K and by a short one (S-Pbca) below $T_S$.
The L- and S-Pbca phases are also observed in Ca$_{2-x}$Sr$_{x}$RuO$_4$ for all $x\le 0.2$,
but $T_S$ decreases with increasing $x$;
for $x>0.2$ the system becomes tetragonal (for $x<1.5$: I4$_1$/acd, {\bf c}-axis rotations only).

Because of the layered structure, the $\sim xz,yz$ band-width, $W_{xz/yz}$,
 is about one half of  the $\sim xy$ band-width, $W_{xy}$. 
Due to the structural distortions, the $t_{2g}$ manifold splits into 
non-degenerate crystal-field states. 
Many-body studies of 3-band Hubbard models show that a large difference in bands widths, a crystal-field splitting 
$\Delta$ and a finite Coulomb exchange interaction can affect the nature of the Mott transition 
\cite{werner,liebsch,OSMT}. 
Simple models neglect, however, the actual effects of distortions
on the electronic structure; such  effects could be crucial \cite{eva} to the mechanism
of the MIT. 
On the other hand, approximate treatments of the many-body effects 
\cite{Fang}, or the neglect of the spin-flip and pair-hopping  contribution to
the Coulomb exchange interaction,  
could also lead to wrong conclusions on the origin of the  transition.

In this Letter we address the problem  by means 
of the LDA+DMFT  (local-density approximation + dynamical mean-field theory) approach \cite{lda+dmft}
with a continuous-time quantum Monte Carlo (QMC) solver \cite{ctqmc}.
This method allows us to treat realistically both the material-dependence and 
the many-body effects. 
We show that the  Mott transition occurs because of the L-$\to$S-Pbca structural phase
transition, which is also responsible for $\sim xy$ orbital order (OO).
In the metallic phases we find, with increasing $x$, a progressive transfer of electrons from  $xy$
to $xz/yz$ ($p\le0$); down to $\sim$300~K, we find, however, no orbital-selective Mott transition.

We use the {\it ab-initio} downfolding approach based on the $N$-th Order Muffin-Tin Orbital (NMTO) method
to construct from first-principles material-specific Wannier functions \cite{eva} which span the $t_{2g}$ bands,
and the corresponding, material-specific,  three-band Hubbard models,
with full local Coulomb interaction \cite{kanamori}
\begin{eqnarray}\label{model} 
\nonumber
H\!\!&=&-\!\!\!\!\sum_{m,m^\prime,i,i^\prime,\sigma} \!\!
t^{i,i^\prime}_{m m^\prime} c^{{\dagger}}_{i m\sigma} c^{\phantom{\dagger}}_{ i^\prime m^\prime\sigma}
+U\sum_{i\; m}\! n_{i m\uparrow}n_{i m\downarrow} \\
\nonumber
&+& \mathop{\sum_{i\sigma \sigma^\prime}}_{m\neq m^\prime} 
\! (U-2J-J\delta_{\sigma,\sigma^\prime})
  n_{i m \sigma}n_{i m^\prime \sigma^\prime} \\
  &-&\!\! J\mathop{\sum_{i\; m\neq m^\prime}} \!\!
  \left[
  c_{i m\uparrow}^{\dagger}
  \left(
  c_{i m^\prime \uparrow}^{\dagger }c^{\phantom{\dagger}}_{i m \downarrow}
  \!+\! 
  c_{i m \downarrow}^{\dagger }c^{\phantom{\dagger}}_{i m^\prime\uparrow}
  \right)
  c^{\phantom{\dagger}}_{i m^\prime\downarrow}
  \right].
\end{eqnarray}
 Here  $c^{\dagger}_{i m\sigma}$ creates an electron with spin $\sigma$ in the Wannier state $m$ 
 at site $i$, and $n_{m\sigma}=c^{\dagger}_{i m\sigma}c^{\phantom{\dagger}}_{i m\sigma}$.
 The  {\it ab-initio}  parameters  $t^{i,i^\prime}_{m m^\prime}$ (Table \ref{hops}) are the crystal-field splittings ($i=i^\prime$)
 and hopping integrals ($i\ne i^\prime$).
 $U$ and $J$ are the direct and exchange screened Coulomb interaction.
We use $U=3.1$~eV and $J=0.7$~eV, in line with experimental \cite{Yokoya,xas,Jung} 
and theoretical  \cite{pchelkina} estimates.
The last row in (\ref{model}) describes the spin-flip and pair-hopping Coulomb terms.
We solve  (\ref{model}) by DMFT \cite{dmft} with a
weak-coupling continuous-time QMC \cite{ctqmc} solver. 
We retain the full self-energy matrix in orbital space, $\Sigma_{m,m^\prime}$ \cite{eva}. 
Our calculations yield the Green function matrix on the imaginary axis; we obtain the spectral matrix 
on the real axis by using a stochastic approach  \cite{mishchenko}.
For limit cases (no spin-flip and pair-hopping terms) we perform comparative calculations with an alternative LDA+DMFT scheme, based on the projection of LDA Bloch states 
(obtained via the projector-augmented wave method \cite{paw}, Vienna Ab-Initio Simulation 
Package \cite{vasp}) to local orbitals \cite{amadon08} and  a Hirsch-Fye QMC solver \cite{hfqmc}. 
The parameters obtained with the two methods are very similar.

\begin{figure}
\center
\rotatebox {0}{\includegraphics [width=0.43\textwidth]{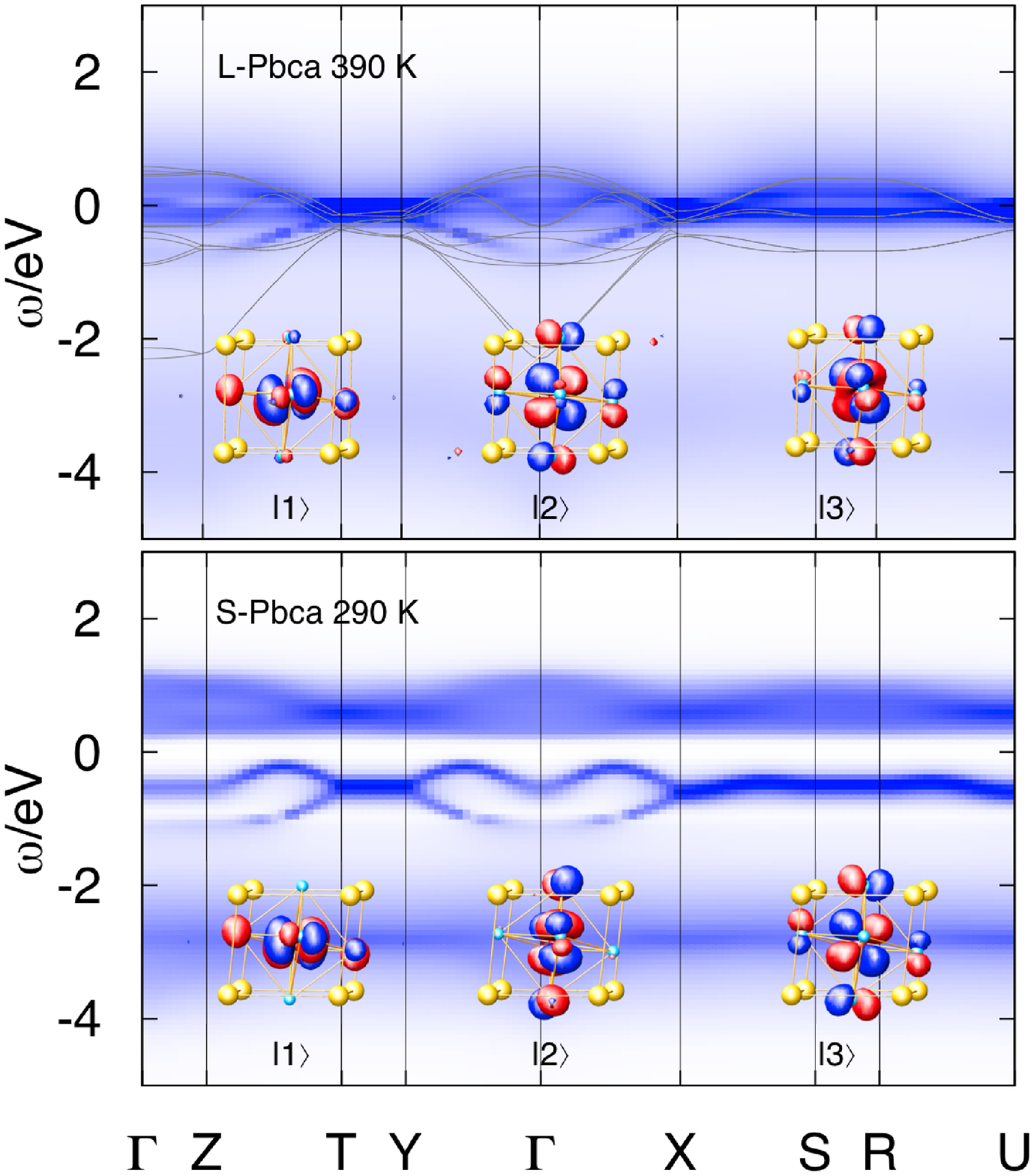}}
\caption{ (Color online) \label{mitdmft}
The Mott transition: Correlated bands for  L- (top) and S-Pbca (bottom), LDA bands (L-Pbca, solid lines), and
crystal-field orbitals for site 1 (see Table \ref{hops}). Positive (negative) lobes are red (blue). 
S-Pbca: $|1\rangle$ is full,  $|2\rangle$ and $|3\rangle$  half-filled;
the valence band has character $|1\rangle$.
}
\end{figure}

The nearest-neighbor hopping integrals $t_{xy,xy}^{i,i+{\bf x}}$, $t_{xz,xz}^{i,i+{\bf x}}$, $t_{xy,xy}^{i,i+{\bf y}}$ progressively decrease going from the ideal tetragonal Sr$_2$RuO$_4$ (see Ref. \cite{igor}) to the L-Pbca and then the S-Pbca structure of Ca$_2$RuO$_4$; correspondingly, the band-width decreases from 2.8~eV (L-Pbca) to 2.5~eV (S-Pbca).
The cause is the increase in tilting 
of the RuO$_6$ octahedra and deformation of the Ca cage, via   Ru-O but also Ca-Ru and Ca-O covalency \cite{eva};
for similar reasons \cite{eva}, the crystal-field splitting increases from $\sim$100~meV (L-Pbca)
to $\sim$300~meV (S-Pbca).
The crystal-field orbitals are displayed in Fig.~\ref{mitdmft}.
For the L-Pbca phase our LDA calculations yield  $p_{\rm \: LDA}\sim 0$.
This may appear surprising, since at 2/3 filling due to the 
difference in band-width ($W_{xz/yz}\sim 1.5$~eV and $W_{xy}\sim 2.8$~eV) one might expect
$p < 0$. Such effect, however, is cancelled by the crystal-field splitting of about 100 meV: since the lowest energy state is $|1\rangle=0.932| xy\rangle+0.259|yz\rangle +0.255 |xz\rangle$ (see Fig.~\ref{mitdmft} and Table \ref{hops}), i.e. close to $|xy\rangle$, neglecting the difference in band-width, the crystal-field 
splitting favors $p > 0$. 
For the S-Pbca structure we find $W_{xz/yz}\sim 1.3$~eV and $W_{xy}\sim 2.5$~eV, i.e. the band-widths decrease by about 0.2~eV. The crystal-field splitting is 300~meV, and the lowest energy crystal-field orbital, $|1\rangle$, is basically identical to $|xy\rangle$.
The effect of the crystal-field is stronger than in the L-Pbca structure leading to $p_{\rm \: LDA}\sim 0.37$.

\begingroup
\squeezetable
\begin{table}
\textbf{L-Pbca Ca$_2$RuO$_4$}
\smallskip
\begin{ruledtabular}
\begin{tabular*}{\columnwidth}{@{\extracolsep{\fill}} cccccccccc}
$lmn$ & $t_{yz,yz}$ & $t_{yz,xz}$ & $t_{yz,xy}$ &  $t_{xz,yz}$ & $t_{xz,xz}$ & $t_{xz,xy}$ & $t_{xy,xy}$ \\
\hline
100   &   10 & -88 & -6&  75   & 242 & -50 & 230     \\
010   & 242 & -88 & -35& 75   & 10   & 26  & 230\\
\end{tabular*}
\end{ruledtabular}
\[
\Delta E_{\alpha,1}=\left(\begin{array}{r}0\\103\\121\end{array}\right),
\left(\begin{array}{c}
		|1\rangle\\
		|2\rangle\\
		|3\rangle
		\end{array}\right)=
		\left(
		\begin{array}{rrr}
		0.259 & 0.255 & 0.932 \\
		0.785 & -0.618 & -0.050 \\
		0.563 & 0.744 & -0.360
		\end{array}	
		\right)\left(\begin{array}{c}
		|yz\rangle\\
		|xz\rangle\\
		|xy\rangle
		\end{array}\right)
\]
\textbf{S-Pbca Ca$_2$RuO$_4$}
\smallskip
\begin{ruledtabular}
\begin{tabular*}{\columnwidth}{@{\extracolsep{\fill}} cccccccccc}
$lmn$ & $t_{yz,yz}$ & $t_{yz,xz}$ & $t_{yz,xy}$ &  $t_{xz,yz}$ &  $t_{xz,xz}$ & $t_{xz,xy}$ & $t_{xy,xy}$ \\
\hline
100   & 11& -61 &  -25 & 38& 123  &  -47 & 205     \\
010   & 123 & -61 & -34 & 38 & 11 &  45 & 205\\
\end{tabular*}
\end{ruledtabular}
\[
\Delta E_{\alpha,1}=\left(\begin{array}{r}0\\308\\341\end{array}\right),
\left(\begin{array}{c}
		|1\rangle\\
		|2\rangle\\
		|3\rangle
		\end{array}\right)=
		\left(
		\begin{array}{rrr}
		0.246 & -0.009 & 0.969 \\
		0.420 & 0.903 & -0.098 \\
		0.874 & -0.430 & -0.227
		\end{array}	
		\right)\left(\begin{array}{c}
		|yz\rangle\\
		|xz\rangle\\
		|xy\rangle
		\end{array}\right)
\]

\textbf{L-Pbca Ca$_{1.8}$Sr$_{0.2}$RuO$_4$}
\smallskip
\begin{ruledtabular}
\begin{tabular*}{\columnwidth}{@{\extracolsep{\fill}} cccccccccc}
$lmn$ & $t_{yz,yz}$ & $t_{yz,xz}$ & $t_{yz,xy}$ &  $t_{xz,yz}$ & $t_{xz,xz}$ & $t_{xz,xy}$ & $t_{xy,xy}$ \\
\hline
100   &   9 & -90 & -6&  87   &275 & -46 & 242     \\
010   & 275 & -90 & -31& 87   &9 &   24       &242\\
\end{tabular*}
\end{ruledtabular}

\caption{\label{hops} Ca$_{2-x}$Sr$_x$RuO$_4$ ($x=0,0.2$): Hopping integrals $t^{ii^\prime}_{m,m^\prime}$/meV
between sites $i^\prime=1$ and $i \sim l{\bf x}+m{\bf y}+n{\bf z}$, and ($x=0$) crystal-field splitting $\Delta E_{\alpha,1}$/meV=$\varepsilon_\alpha-\varepsilon_1$
($\alpha=1, 2 ,3 $)  and orbitals  at site 1.
Orbitals and hopping integrals for sites 2, 3 and 
4 can be obtained  using symmetries (Fig.~\ref{crys-str}).
}
\end{table}
\endgroup

\begin{figure}
\center
\rotatebox {0}{\includegraphics [width=0.45\textwidth]{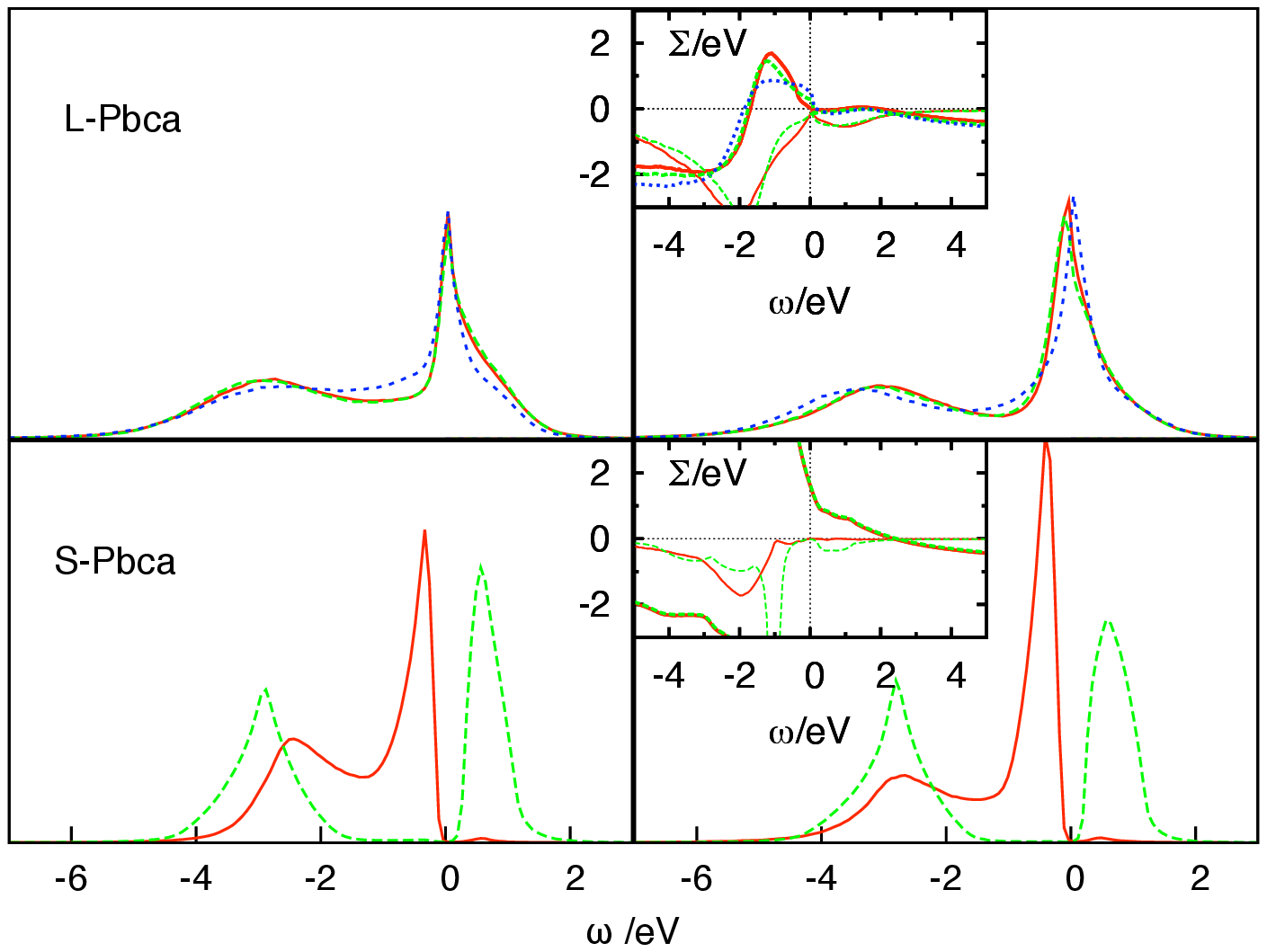}
}
\caption{(Color online) \label{fig1a}
Spectral matrix. Solid lines: $A_{1,1}$. Dashed lines: $A_{2,2}$ and $A_{3,3}$.
Left: density-density terms only. Right: rotationally invariant Coulomb vertex.
First row: L-Pbca, $T$=390~K.
Second row: S-Pbca, $T$=290~K.
Inset: Real (thick lines) and imaginary (thin lines) self-energies. 
}
\end{figure}

The LDA+DMFT solution of Hamiltonian (\ref{model}) yields the following results:
For the L-Pbca structure we find a metallic solution down to very low temperatures; the 
orbital polarization is $p\sim 0$ at 390~K, i.e. close to the LDA value.
The self-energy (Fig.~\ref{fig1a}) exhibits a narrow
Fermi-liquid regime with kinks \cite{kinks}; 
a (lower) estimate of the effective masses of the quasi-particles, obtained from 
the slope of ${\rm Im} \Sigma_{m,m} (i\omega_n)$ at the first Matsubara frequency, is $m^*/m\sim 5.0$ ($xy$) 
and $m^*/m\sim 4.2$ ($xz,yz$). 
We find similar behavior in Sr$_2$RuO$_4$; in Ca$_2$RuO$_4$ the mass enhancement is larger, because of the narrower
band-width. Lowering the temperature down to 290~K turns the system into a ferromagnet with (almost) half-metallic
behavior and  {$p\sim-0.05$ (the occupation of $|xy\rangle$ slightly decreases)}.
The correlated bands  for Ca$_2$RuO$_4$  are shown in Fig. \ref{mitdmft}. Along $\Gamma$X 
we find dispersive bands, while along  XS  the bands become almost flat.

For the S-Pbca structure the situation is completely different. We find a MIT
between 390~K and 290~K, in very good agreement with the experimental $360$~K \cite{mit}. 
At 580~K the spectral function exhibits a pseudogap; $n_{xy}\sim 1.9$ and $n_{xz/yz}\sim 1.1$, 
and the polarization is already as large as $p\sim 0.8$. At 290~K the gap is open and about $0.2$~eV
wide (see Fig.\ref{fig1a}), while
$p\sim 1$. The most occupied state is basically identical to the 
LDA lowest energy crystal-field orbital (Table (\ref{hops}));  the orbital order
is close to $xy$ ferro-order, with a small antiferro component. 
LDA+$U$ calculations for the anti-ferromagnetic phase yield an OO consistent with our results for the paramagnetic phase \cite{OSMT,Fang}.
Such orbital order in the S-Pbca phase and none in the L-Pbca phase
is in line with the evolution, across the MIT, of 
the O 1s x-rays absorption (XAS) spectra \cite{xas,xas2} with increasing light incidence angle, $\theta$.
Our  spectral functions are also consistent with photoemission \cite{xas2,pes}:
the quasi-particle peak in the high-temperature phase, the Hubbard band at $-0.5$~eV and the
multiplets at -3 eV all correspond to features in the experimental spectra.
The experimental peak centered at $\sim -1.5$ eV in \cite{xas2,pes}, 
absent in our $t_{2g}$ spectral functions, likely includes a sizable contribution of the O bands, which start around those energies.
The small gap (S-Pbca) is  in excellent agreement with electrical resistivity \cite{mit,ca2x} and
optical conductivity \cite{gap} data. 

In Fig.~\ref{fig1a} we investigate the role played by the spin-flip 
and pair hopping Coulomb terms by comparing
results with and without those terms. We find that
the effective masses decrease when the full Coulomb interaction is considered; 
this happens because the average $U$ decreases and 
the degeneracy of the low energy $d^3$ and $d^4$ multiplets increases. In the insulating phase, 
such difference in multiplet degeneracy appears very clearly as the enhancement of the $d^4 \to d^3$ peak 
around $\sim -0.5$~eV. The differences in orbital polarization  
are  small.

Our results show that the Mott transition in Ca$_2$RuO$_4$ is  
driven by the L$\to$S-Pbca structural transition at $T_S$ and occurs simultaneously with orbital order. 
For the L-Pbca structure we find no OO for temperatures well below T$_S$; OO
follows, rather than produces, the change in structure \cite{oo}.
These findings are in very good agreement with resonant x-rays scattering interference data \cite{rxr},
which show that the $xy$ OO disappears close to the the MIT.
They are also in line with the fact \cite{pressure} that a tiny pressure ($\sim$ 0.5 GPa) is sufficient to make the system metallic  at 290~K  
via the S- to L-Pbca  transition \cite{pressure}. 

Can this scenario be extended to Ca$_{2-x}$Sr$_{x}$RuO$_4$ ($x\le0.2$)?
Let us examine the limit system $x=0.2$, for which the L-Pbca phase persists down to 10~K.
Neglecting the chemical effects of the Ca $\to$ Sr substitution \cite{sr} and disorder effects, for the 10~K structure 
we find crystal-field splittings of 81 and 110 meV, slightly ($\sim 0.1$~eV) 
broader $t_{2g}$ bands than for $x=0$, and $p_{\rm \: LDA}\sim -0.03$.  The main difference with the L-Pbca structure of Ca$_2$RuO$_4$  is in the crystal-field orbitals 
(e.g. $\langle xy|1\rangle\sim\langle xy |2\rangle\sim 0.66$), and can be ascribed to 
the differences in octahedra tilting and rotation, and corresponding distortions of the cation cage.
All these effects stabilize the metallic solution.
With LDA+DMFT (390-290~K) we find three metallic bands, in line with ARPES results from Ref.~\cite{Shimo}, with  
$m^*/m\sim3.7$ ($xz$), 4.4 ($yz$), 5.6 ($xy$); $p\sim -0.14$. 
While the details slightly differ, depending on $x$ \cite{x}, we conclude  that for $x\le 0.2$ the temperature-induced Mott transition is {mostly} driven by the change in structure L-Pbca $\to$ S-Pbca.  We find no OSMT down to 290~K.

What happens for $x>0.2$? For the $x=0.5$  structure \cite{sr} the crystal-field states are $|1\rangle=|xy\rangle$, and $|xz\rangle$, $|yz\rangle$, the crystal-field splitting is small, $p_{\rm \: LDA}=-0.02$,  $W_{xy}\sim2.7$~eV, $W_{xz/yz}\sim1.6$~eV. LDA+DMFT at 390~K yields again a metallic solution with
$m^*/m\sim4.0$ ($xz/yz$) and 5.6 ($xy$), and three metallic bands, in agreement with ARPES \cite{Wang}; $p\sim-0.15$ at 390-290~K.

Thus, outside the S-Pbca phase we always find a metal, in line with transport and optical conductivity data \cite{str2,ca2x,optics};  with increasing $x$, $n_{xz}+n_{yz}$ increases, in line with XAS \cite{xas3}; $p\sim 0$ or slightly negative, approaching the $p= -1/2$ of the OSMT scenario \cite{OSMT}; we find, however, 
no OSMT down to 290~K.

In conclusion, we have studied the origin of the metal-insulator transition in Ca$_{2}$RuO$_4$.
We find that it is driven by the structural L-$\to$S-Pbca transition. 
Two mechanisms compete; while a small $W_{xz/yz}/W_{xy}$ band width ratio enhances the occupation of
the $xz/yz$ orbitals ($p<0$) and could lead to a orbital-selective Mott transition  with $p=-1/2$ \cite{OSMT}, 
a large crystal-field splitting $\Delta_{3,1}$, with  $|1\rangle\sim |xy\rangle$ as the lowest energy state, 
favors $xy$ orbital order and $p>0$, as in Ref. \cite{liebsch}. 
In the $x=0$ L-Pbca structure the two effects compensate: $\Delta_{3,1}/W_{xy}\sim 0.04$, $\langle 1|xy\rangle\sim 0.93$ and $W_{xz/yz}/W_{xy}\sim 0.54$.  We find a metallic solution above and well below $T_{\rm{MIT}}\sim 360~$K, with orbital polarization $p\sim0$ (no orbital order)  at $T\sim390$~K. At low temperatures the system becomes a ferromagnetic metal, in line with  (moderate) pressure studies \cite{pressure}.
In the $x=0$ S-Pbca structure $\Delta_{3,1}/W_{xy}\sim 0.13$, sizeably larger than for the L-Pbca structure, 
$\langle 1|xy\rangle\sim 0.97$,  and $W_{xz/yz}/W_{xy} \sim 0.52$; 
the system becomes insulating around $T_{\rm{MIT}}$ and $p\sim 1$ ($\sim xy$ ferro orbital order),
in excellent agreement with experiments; orbital order follows, rather than drives, the transition.   
Our results indicate that this scenario can be extended to
Ca$_{2-x}$Sr$_x$RuO$_4$ for all  $x\le0.2$. 
Finally, for the metallic $x\le0.5$ phases we find that, differently than in the crystal-field scenario \cite{liebsch},
$p \sim 0$ or negative, slowly approaching the $-1/2$ of Ref.~\cite{OSMT} with increasing $x$, but, down to $\sim$~300~K,  we find no orbital-selective Mott transition.

Calculations were performed on the J\"ulich BlueGene/P. We thank
A.~Hendricks, T.Koethe, and H.~Tjeng for sharing with us their data 
 before publication.


\begin{thebibliography}{99}
\bibitem{sr}
The effect of the full substitution Ca $\to$ Sr is the following: for $x=0.2$ 
both crystal-field splitting (76 and 147 meV) and $t_{2g}$ band-width increase slightly, 
$\langle1|xy\rangle\sim 0.89$, $p_{\rm LDA}\sim0.16$. 
For $x=0.5$ the crystal-field splitting increases to $\sim 80$~meV, with $|1\rangle=| xy\rangle$,
and $p_{\rm LDA}=0.25$.
\bibitem{x} 
For $x=0.1$ LDA+DMFT (projection scheme) yields a metal for L-Pbca ($p\sim 0.03$ at $580~$K) and a deep pseudogap for S-Pbca ($p\sim 0.85$ at $390~$K). For $x=0.15$ (L-Pbca) we find a metal at $390~$K with $p\sim -0.07$.
\end{thebibliography}
\end{document}